\documentclass[10pt,conference]{IEEEtran}

\usepackage{amsmath,amssymb,amsfonts,mathtools}
\usepackage{bm,microtype,flushend}
\usepackage{algorithm}
\usepackage{algpseudocode}
\usepackage{booktabs}
\usepackage{graphicx}
\graphicspath{{fig/}}
\usepackage{placeins}

\usepackage{cite}
\usepackage{url}
\usepackage[hidelinks]{hyperref}
\hypersetup{pdftitle={Model-Free Federated LQR with Scalar Communication and Personalized Rollout Allocation},
 pdfauthor={Mohammadreza Rostami, Shahriar Talebi, Solmaz S. Kia}}
\usepackage{amsthm}

\usepackage{tikz}
\usetikzlibrary{positioning,arrows.meta,fit,backgrounds,calc}

\newcommand{\R}{\mathbb{R}}
\newcommand{\E}{\mathbb{E}}

\newcommand{\fro}[1]{\left\|#1\right\|_{\mathrm F}}
\newcommand{\tr}{\operatorname{Tr}}
\newcommand{\diag}{\operatorname{diag}}
\newcommand{\cS}{\mathcal{S}}
\newcommand{\cH}{\mathcal H}
\newcommand{\cV}{\mathcal V}

\newcommand{\inner}[2]{\langle #1,#2\rangle_{\rm F}}

\newtheorem{theorem}{Theorem}
\newtheorem{lemma}{Lemma}
 \newtheorem{proposition}{Proposition}
   \newtheorem{corollary}{Corollary}
\newtheorem{remark}{Remark}
\newtheorem{assumption}{Assumption}
\newtheorem{problem}{Problem}
\allowdisplaybreaks[2]
\IEEEoverridecommandlockouts

\title{%Personalized Rollout Allocation in Federated Learning for Stochastic LQR Problems
Personalized Rollout Allocation for Federated LQR Policy Optimization under Heterogeneous Process~Noise}

\author{Mohammadreza Rostami, Shahriar Talebi, \emph{Member, IEEE} and Solmaz S. Kia, \emph{Senior Member, IEEE}
\thanks{This work is partially supported by the 2026 UCLA--UCI PI grant awarded by the Samueli Foundation. \emph{Corresponding author}: M. Rostami }
\thanks{M. Rostami and S. Kia are with the University of California, Irvine, CA, US, and S. Talebi is with the University of California, Los Angeles, CA, US. Emails: {\em\small mrostam2@uci.edu}, {\em\small s.talebi@ucla.edu} and {\em\small solmaz@uci.edu}}
}

\begin{document}
\maketitle
\begin{abstract} We study federated model-free learning of a common linear quadratic regulator (LQR) gain under a limited uplink communication budget, with clients that share dynamics and costs but experience different process-noise levels. Conventionally, clients transmit gradient matrices whose dimension equals the number of controller parameters. Exploiting the structure of two-point zeroth-order estimates, we consider an implementation in which clients transmit only one scalar cost-difference coefficient per rollout pair, while the server regenerates the associated perturbation directions from shared seeds. Communication thus scales with the number of rollout pairs rather than the controller dimension, turning the uplink budget into a rollout budget. Because noise heterogeneity produces gradient estimates of widely differing variance and every client carries equal weight in the fleet objective, uniform rollout allocation allows noisy clients to dominate the aggregated update and increases the probability of destabilizing updates. We propose a rollout-allocation scheme that assigns rollout pairs in proportion to the standard deviation of each client's estimates and show that it minimizes the variance of the aggregated update. The scheme is implemented exactly through a greedy integer allocation rule whose weights are estimated from the communicated scalars. We further prove that a single allocation factor quantifies its improvement over uniform allocation in a one-round instability bound, the certified step size, and the error floor of a joint stability and convergence guarantee. Simulations show that, under the same communication budget, personalized allocation reduces the one-step instability rate from $37\%$ to $0.3\%$. \end{abstract}


\section{Introduction}\label{sec:intro}

In federated linear quadratic regulator (LQR) learning, a server
aggregates policy-gradient information computed from trajectory
rollouts performed by distributed clients and uses the aggregate to
update a shared control gain in an iterative, model-free manner. This
paper focuses on settings in which uplink communication is severely
limited, for example by bandwidth constraints or by the energy
available on battery-powered platforms, and in which clients share
dynamics and costs but experience different process-noise levels.

To reduce uplink traffic, we exploit the structure of two-point
zeroth-order gradient
estimates~\cite{nesterov2017random,duchi2015optimal,shamir2017optimal}.
Each estimate consists of a scalar cost difference obtained from a pair
of rollouts and a random perturbation direction. Since the perturbation
directions can be regenerated by the server from shared random seeds,
clients need only transmit the scalar coefficients. This reduces
communication whenever a client performs fewer rollout pairs than there
are controller parameters. More importantly, because each transmitted
scalar is tied to a particular perturbation direction, the coefficients
cannot be aggregated locally before communication. As a result, uplink
communication scales with the number of rollout pairs rather than with
the controller dimension, and a fixed communication budget becomes a
budget on the total number of rollout pairs available across the fleet.
In this setting, communication-efficient learning becomes a
resource-allocation problem: how should rollout pairs be distributed
across clients?

The allocation matters because clients experience different
process-noise levels and therefore produce gradient estimates of
different variance. The variance of a two-point estimate combines the
randomness of the perturbation direction with the noise of the measured
costs~\cite{nesterov2017random,ghadimi2013stochastic,duchi2015optimal},
and in LQR the latter is driven by the process noise acting on each
finite-length
rollout~\cite{malik2019derivative,mohammadi2022convergence,song2025stochastic}.
Since all clients contribute equally to the fleet objective, noisy
clients cannot simply be down-weighted. Under uniform allocation, a
small number of high-noise clients can dominate the variance of the
aggregated update. When only a limited number of rollout pairs can be
collected in each round, the resulting sampling error does not merely
slow learning but can drive the updated gain, or the perturbed gains
queried around it, outside the stabilizing
set~\cite{fazel2018global,malik2019derivative,song2025stochastic}.
Consequently, rollout allocation should be guided by the reduction in
aggregate uncertainty achieved by additional rollout pairs rather than
by local data quality alone. We refer to the reduction in the variance
of the aggregated update achieved by one additional rollout pair from a
client as the \emph{effectiveness} of that client; it grows with the
variance of the client's estimates and diminishes as more rollout pairs
are assigned to it.

\textit{Related work.} LQR is a canonical benchmark for learning-based
control~\cite{recht2019tour}. Despite the nonconvexity of the set of
stabilizing gains, policy-gradient methods admit global convergence
guarantees~\cite{fazel2018global,bu2019lqr,hu2023toward}, and these
guarantees extend to settings with additive or multiplicative process
noise~\cite{gravell2021learning,hambly2021policy,mohammadi2022convergence,song2025stochastic}.
Model-free methods avoid system identification~\cite{dean2020sample} at
the cost of trajectory rollouts, whose sample complexity and estimation
error have been analyzed
in~\cite{tu2019gap,malik2019derivative,song2025stochastic}. Two-point
estimators further connect model-free control with zeroth-order
optimization~\cite{nesterov2017random,ghadimi2013stochastic,duchi2015optimal,shamir2017optimal}.
Federated formulations of controller learning have addressed
heterogeneous dynamics~\cite{wang2023federated},
meta-learning~\cite{toso2024meta}, variance
reduction~\cite{toso2024oracle}, decentralized
control~\cite{li2022distributed}, and federated reinforcement
learning~\cite{khodadadian2022federated}. In these works, heterogeneity
is primarily modeled through differences in dynamics or tasks, whereas
heterogeneity in the quality of locally collected data has received
comparatively little attention.

Communication bottlenecks in federated learning have motivated
compressed and quantized update
representations~\cite{mcmahan2017communication,alistarh2017qsgd}. The
scalar-and-seed structure of random-direction estimators has been
exploited in federated zeroth-order
optimization~\cite{fang2022communication,qin2024federated}, and a
scalar communication scheme was recently proposed for federated
model-free LQR~\cite{rostami2026scalar, rostami2026scalar_federated, rostami2025fedmpdd}. These methods, however,
typically allocate the same number of samples to every client.
Nonuniform sampling and importance sampling are known to reduce
estimator
variance~\cite{chen2022optimal,needell2014stochastic,zhao2015stochastic},
and the resulting square-root rule coincides with Neyman allocation in
stratified sampling~\cite{neyman1934two}. When clients contribute
unequal amounts of local work, their updates must be normalized to
avoid biasing the aggregate objective~\cite{li2020federated,wang2020tackling}.
None of these works, however, relates rollout allocation to closed-loop
stability or to the convergence properties of learning-based control.

This paper addresses this gap by developing a communication-aware
federated framework for model-free LQR in which clients communicate
scalar coefficients and rollout pairs are allocated according to the
effectiveness of each client.

\textit{Contributions.} The main contributions are as follows.
\begin{itemize}
\item We formulate rollout allocation under a fixed scalar
communication budget and derive the variance-optimal allocation, which
assigns rollout pairs in proportion to the standard deviation of each
client's coefficients. We further develop an exact integer
implementation that allocates each rollout pair to the currently most
effective client while guaranteeing minimum participation, and show
that the required weights can be estimated from the communicated
coefficients without additional communication.
\item We establish a one-round stability certificate that bounds the
probability of destabilizing updates and destabilizing queries. A
single allocation factor $\Gamma_k(S)$ divides both this bound and the
communication budget required to certify stability, and enlarges the
certified step size by up to $\sqrt{\Gamma_k(S)}$.
\item We derive a joint stability and convergence guarantee for the
fleet-averaged LQR objective and show that the same allocation factor
reduces its sampling-error floor.
\end{itemize}
Simulations with fleets of ten and twenty clients support the analysis.
Under the same communication budget, personalized allocation reduces
the one-step instability rate from $37\%$ to $0.3\%$ and sustains
learning at step sizes for which uniform allocation fails.

The remainder of the paper is organized as follows.
Section~\ref{sec:problem} presents the system model and the scalar
communication scheme, Section~\ref{sec:allocation} derives the
allocation, Section~\ref{sec:stability} establishes the stability and
convergence guarantees, and Section~\ref{sec:numerics} reports the
simulations.
\section{Model-Free Federated LQR with Scalar Communication}
\label{sec:problem}
We consider $M$ clients, each operating an identical copy of an unknown
linear system with dynamics matrices $A$ and $B$ and quadratic cost
matrices $Q\succ0$ and $R\succ0$. Under the linear state feedback
$u_t^{(i)}=-Kx_t^{(i)}$, with $K\in\R^{n_u\times n_x}$, the state of
client $i$ evolves as
\begin{equation}
 x_{t+1}^{(i)}=A_Kx_t^{(i)}+w_t^{(i)},\qquad A_K=A-BK,
 \label{eq:closed-loop}
\end{equation}
and we write $D=n_un_x$ for the number of controller parameters. The
set of stabilizing gains is $\cS=\{K:\rho(A_K)<1\}$, and a stabilizing
initial gain $K_0\in\cS$ is assumed to be available. The clients differ
only in the statistics of their process noise, as specified next.

\begin{assumption}[Process noise]\label{ass:noise}
For each client $i$, the disturbances $\{w_t^{(i)}\}$ are independent
and identically distributed with zero mean, covariance $W_i\succeq0$,
and finite fourth moment, and they are independent across rollouts and
across clients. Initial states have finite fourth moment and are
independent of subsequent disturbances.
\end{assumption}

Because the process noise acts persistently, the performance of a gain
$K$ on client $i$ is measured by the average cost
\begin{equation}
J_i(K)=\lim_{T\to\infty}\frac1T\,\E\left[\sum\nolimits_{t=0}^{T-1}
 x_t^{(i)\top}(Q+K^\top RK)\,x_t^{(i)}\right],
 \label{eq:global-cost}
\end{equation}
and the fleet seeks a common gain that minimizes
$J_{\rm avg}(K)=\frac{1}{M}\sum_{i=1}^MJ_i(K)$ over $\cS$. For every
$K\in\cS$, the average cost admits the representation
$J_i(K)=\tr(P_KW_i)$, where $P_K$ is the unique solution of the Lyapunov
equation $P_K=Q+K^\top RK+A_K^\top P_KA_K$~\cite{song2025stochastic}.
Consequently, all clients share the same stabilizing set and the same
matrix $P_K$, and their objectives differ only through the noise
covariances $W_i$. This representation involves second-order statistics
only and does not require Gaussian disturbances.

Since the model is unknown, the clients estimate gradients from the
observed costs of executed trajectories. Let $\widehat J_i(K)$ denote
the cost measured from one rollout of gain $K$ on client $i$, with a
rollout length, initial condition, and stage-cost averaging that are
common to all clients. At round $k$, client $i$ draws a direction
$U_{i,k,s}$ uniformly from the unit Frobenius sphere in
$\R^{n_u\times n_x}$ and executes two independent rollouts at the
perturbed gains $K_k\pm rU_{i,k,s}$, where $r>0$ is a common smoothing
radius. We call these two rollouts a \emph{rollout pair} and combine
their costs into the scalar coefficient
\begin{equation}
 d_{i,k,s}=\frac{\widehat J_i^+(K_k+rU_{i,k,s})
                   -\widehat J_i^-(K_k-rU_{i,k,s})}{2r}\in\mathbb{R},
 \label{eq:scalar-dd}
\end{equation}
which defines the gradient estimate $Z_{i,k,s}=D\,d_{i,k,s}U_{i,k,s}$.
Client $i$ is assigned $m_{i,k}$ rollout pairs at round $k$ and thus
produces $m_{i,k}$ coefficients. The estimate $Z_{i,k,s}$ carries a
smoothing bias due to the finite radius~\cite{shamir2017optimal} and a
variance due to the random direction and the trajectory noise, which
are characterized in Sections~\ref{sec:allocation}
and~\ref{sec:stability}; the perturbed gains must themselves be
stabilizing, and conditions that guarantee this are also given in
Section~\ref{sec:stability}.

\paragraph*{Scalar communication}
The estimate $Z_{i,k,s}$ is the product of a data-dependent scalar
$d_{i,k,s}$ and a data-independent direction $U_{i,k,s}$. Before
learning begins, the server and the clients agree on a seed $q_0$, and
each direction is generated deterministically from the client, round,
and rollout indices, e.g., from $q_{i,k}=\mathsf{Hash}(q_0,i,k)$. Client
$i$ therefore transmits only $\{d_{i,k,s}\}_{s=1}^{m_{i,k}}$, and the
server regenerates the directions and reconstructs each $Z_{i,k,s}$
exactly; in the analysis, the directions are modeled as independent
uniform draws. Because each coefficient is meaningful only together
with its own direction, the coefficients cannot be aggregated before
transmission, so the uplink load of a round is $\sum_im_{i,k}$ scalars,
compared with $MD$ entries when each client uploads a locally averaged
gradient matrix. Scalar transmission is thus cheaper for client $i$
whenever $m_{i,k}<D$, and a per-round budget of $S$ transmitted scalars
is a budget of $S$ rollout pairs across the fleet. We exclude the
downlink broadcast and the one-time seed agreement from the budget and
treat the received coefficients as exact.

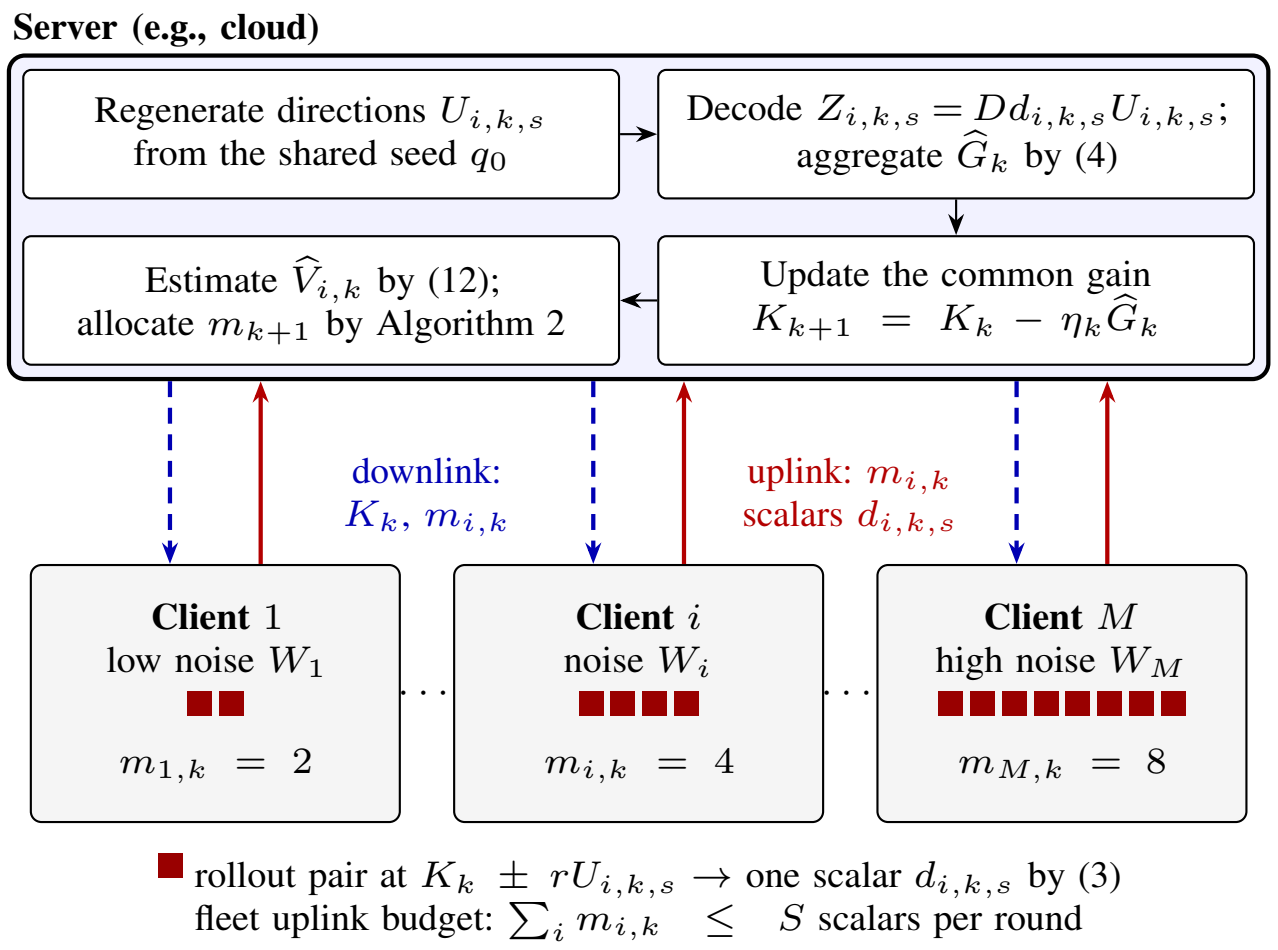
\begin{figure}[t]
\centering
\begin{tikzpicture}[
  font=\scriptsize,
  >={Stealth[length=4pt]},
  blk/.style={draw, rounded corners=1.5pt, fill=white, align=center,
              inner sep=2pt, text width=38mm, minimum height=8.5mm},
  cli/.style={draw, rounded corners=2pt, fill=gray!8, align=center,
              inner sep=2pt, text width=23mm, minimum height=17mm},
  down/.style={->, densely dashed, blue!70!black, thick},
  up/.style={->, red!70!black, thick},
]
\newcommand{\pairs}[1]{\tikz[baseline=-0.6mm]{\foreach \j in {1,...,#1}
  {\fill[red!60!black, sharp corners] (\j*2.1mm,0) rectangle ++(1.6mm,1.6mm);}}}
% server pipeline
\node[blk] (regen) at (19.5mm,0)
  {Regenerate directions $U_{i,k,s}$\\ from the shared seed $q_0$};
\node[blk] (dec) at (61.5mm,0)
  {Decode $Z_{i,k,s}\!=\!Dd_{i,k,s}U_{i,k,s}$;\\ aggregate $\widehat G_k$ by~\eqref{eq:global-ZO}};
\node[blk] (upd) at (61.5mm,-11mm)
  {Update the common gain\\ $K_{k+1}=K_k-\eta_k\widehat G_k$};
\node[blk] (alloc) at (19.5mm,-11mm)
  {Estimate $\widehat V_{i,k}$ by~\eqref{eq:sample-variance};\\ allocate $m_{k+1}$ by Algorithm~\ref{alg:alloc}};
\draw[->] (regen) -- (dec);
\draw[->] (dec) -- (upd);
\draw[->] (upd) -- (alloc);
\begin{scope}[on background layer]
  \node[draw, thick, rounded corners=3pt, fill=blue!5,
        fit=(regen)(dec)(upd)(alloc), inner sep=2.5pt,
        label={[anchor=south west, inner sep=1pt, font=\scriptsize\bfseries]
               north west:Server (e.g., cloud)}] (srv) {};
\end{scope}
% clients
\node[cli] (c1) at (12.5mm,-37mm)
  {\textbf{Client $1$}\\ low noise $W_1$\\[2pt] \pairs{2}\\[1pt] $m_{1,k}=2$};
\node[cli] (ci) at (40.5mm,-37mm)
  {\textbf{Client $i$}\\ noise $W_i$\\[2pt] \pairs{4}\\[1pt] $m_{i,k}=4$};
\node[cli] (cM) at (68.5mm,-37mm)
  {\textbf{Client $M$}\\ high noise $W_M$\\[2pt] \pairs{8}\\[1pt] $m_{M,k}=8$};
\node at ($(c1.east)!0.5!(ci.west)$) {$\cdots$};
\node at ($(ci.east)!0.5!(cM.west)$) {$\cdots$};
% links
\foreach \c in {c1,ci,cM}{
  \draw[down] ([xshift=-3mm]srv.south -| \c.north) -- ([xshift=-3mm]\c.north);
  \draw[up]   ([xshift=3mm]\c.north) -- ([xshift=3mm]srv.south -| \c.north);
}
\node[blue!70!black, align=center] at (26.5mm,-24mm)
  {downlink:\\ $K_k$, $m_{i,k}$};
\node[red!70!black, align=center] at (54.5mm,-24mm)
  {uplink: $m_{i,k}$\\ scalars $d_{i,k,s}$};
% legend
\node[align=center, text width=82mm] at (40.5mm,-50.5mm)
  {\pairs{1}\ rollout pair at $K_k\pm rU_{i,k,s}$ $\to$ one scalar $d_{i,k,s}$ by~\eqref{eq:scalar-dd}\\
   fleet uplink budget: $\sum_{i} m_{i,k}\le S$ scalars per round};
\end{tikzpicture}
\caption{One round of the protocol. The server broadcasts the gain $K_k$
and the rollout counts (dashed). Each client executes its assigned
rollout pairs on its own system and uploads one scalar coefficient per
pair (solid), subject to a fleet budget of $S$ scalars per round. The
server regenerates the perturbation directions from the shared seed,
reconstructs the gradient estimate, updates the gain, and reallocates
the next round's counts from the variance of the received coefficients,
so that clients with noisier measurements are assigned more rollout
pairs.}
\label{fig:protocol}
\end{figure}

\paragraph*{Aggregation and protocol}
The server averages the estimates of each client and then across
clients,
\begin{equation}
 \widehat G_{i,k}=\frac1{m_{i,k}}\sum\nolimits_{s=1}^{m_{i,k}} Z_{i,k,s},\qquad
 \widehat G_k=\frac1M\sum\nolimits_{i=1}^M\widehat G_{i,k},
 \label{eq:global-ZO}
\end{equation}
and updates $K_{k+1}=K_k-\eta_k\widehat G_k$ with step size $\eta_k>0$.
Normalizing each local sum by its own count keeps every client's weight
in $J_{\rm avg}$ equal to $1/M$, so the counts do not change the mean of
the update, only its variance. Fig.~\ref{fig:protocol} and
Algorithm~\ref{alg:fedlqr} summarize one round: the server broadcasts
$K_k$ and the counts, each client executes its rollout pairs and uploads
its coefficients, and the server forms the update and computes the next
counts from the sample variances of the received coefficients
(Section~\ref{sec:allocation}). The first counts are obtained from a
balanced calibration batch at $K_0$, charged to the total communication.

\begin{algorithm}[t]
\small
\caption{Federated LQR with personalized rollout counts}
\label{alg:fedlqr}
\begin{algorithmic}[1]
\Require stabilizing $K_0$ with a stabilizing radius-$r$ neighborhood;
number of rounds $N$; radius $r$; step sizes $\{\eta_k\}$; shared seed
$q_0$; budget $S\ge Mm_{\min}$ with $m_{\min}\ge2$; variance floor
$v_{\rm floor}>0$.
\State \textbf{Calibration:} each client uploads $S/M$ coefficients at
$K_0$; the server computes $\widehat V_{i}$ by~\eqref{eq:sample-variance}
and sets $m_{0}\gets\mathsf{Alloc}(S,\{\max(v_{\rm floor},\widehat V_{i})\},m_{\min})$.
\For{$k=0,\ldots,N-1$}
  \State Server broadcasts $K_k$ and the counts $\{m_{i,k}\}_{i=1}^M$.
  \ForAll{clients $i$ in parallel}
    \State Generate $\{U_{i,k,s}\}_{s=1}^{m_{i,k}}$ from the seed $q_{i,k}=\mathsf{Hash}(q_0,i,k)$.
    \State Execute $m_{i,k}$ rollout pairs at $K_k\pm rU_{i,k,s}$; upload $\{d_{i,k,s}\}_{s=1}^{m_{i,k}}$ computed by~\eqref{eq:scalar-dd}.
  \EndFor
  \State Server regenerates $U_{i,k,s}$, forms $Z_{i,k,s}=Dd_{i,k,s}U_{i,k,s}$ and $\widehat G_k$ by~\eqref{eq:global-ZO}, and sets $K_{k+1}=K_k-\eta_k\widehat G_k$.
  \State Server computes $\widehat V_{i,k}$ by~\eqref{eq:sample-variance} and sets
  $m_{k+1}\gets\mathsf{Alloc}(S,\{\max(v_{\rm floor},\widehat V_{i,k})\},m_{\min})$.
\EndFor
\State \Return $K_N$.
\end{algorithmic}
\end{algorithm}

\paragraph*{Variance of the update}
To analyze the effect of the counts, we condition on the history
$\cH_k$ available to the server at the start of round $k$, namely the
coefficients received in all preceding rounds together with the gain
$K_k$, the counts $m_k=\{m_{i,k}\}_{i=1}^M$, and the step size $\eta_k$
determined from them. Conditional on $\cH_k$, the estimates collected
in round $k$ are independent across clients and across rollout pairs,
since each is computed from a fresh direction and fresh trajectory
noise. We denote their conditional mean and variance by
\begin{equation}
 \overline G_{i,k}=\E[Z_{i,k,s}\mid\cH_k],\qquad
 V_{i,k}=\E\big[\fro{Z_{i,k,s}-\overline G_{i,k}}^2\,\big|\,\cH_k\big],
 \label{eq:cond-moments}
\end{equation}
and write $\overline G_k=\frac1M\sum_{i=1}^M\overline G_{i,k}$ for the
conditional mean of the update.

The variance $V_{i,k}$ has two known sources: the randomness of the
perturbation direction, which is of order $D\fro{\nabla J_i(K_k)}^2$
for two-point sphere estimators~\cite{nesterov2017random,shamir2017optimal},
and the noise of the two measured costs, which is amplified by the
factor $D^2/r^2$~\cite{duchi2015optimal,malik2019derivative} and, in
LQR with additive process noise, grows with the noise covariance and
decreases with the rollout
length~\cite{mohammadi2022convergence,song2025stochastic}. We observe
two consequences for the federated setting. First, both sources depend
on the current gain $K_k$, so $V_{i,k}$ changes from round to round.
Second, neither depends on the number of rollout pairs assigned to the
client; the count changes only how many independent copies of
$Z_{i,k,s}$ are averaged. Clients with larger process noise therefore
have larger $V_{i,k}$, and assigning them more rollout pairs is the
only means available to the server of reducing their contribution.

We normalize each local average in~\eqref{eq:global-ZO} by its own
count so that the conditional mean $\overline G_k$ is independent of the
counts. With this choice, conditional independence gives
\begin{equation}
 \cV_k(m):=\E\big[\fro{\widehat G_k-\overline G_k}^2\,\big|\,\cH_k\big]
 =\frac1{M^2}\sum\nolimits_{i=1}^M\frac{V_{i,k}}{m_{i,k}},
 \label{eq:centered-error}
\end{equation}
so the counts affect the update only through its variance. Since
$\overline G_k$ differs from $\nabla J_{\rm avg}(K_k)$ by a bias that
does not depend on the counts, an allocation that minimizes $\cV_k(m)$
also minimizes the mean-squared error with respect to the true
gradient, and we can design the allocation without knowledge of the
bias. We note that under uniform counts, \eqref{eq:centered-error} is
dominated by the clients with the largest $V_{i,k}$. This observation
leads us to pose the following problem.

\begin{problem}\label{prob:allocation}
Given a per-round budget of $S$ transmitted scalars and a minimum
participation level $m_{\min}$, determine the rollout counts
$\{m_{i,k}\}_{i=1}^M$ with $\sum_im_{i,k}\le S$ and $m_{i,k}\ge m_{\min}$
that minimize $\cV_k(m)$ in~\eqref{eq:centered-error}, and characterize
the effect of this allocation, relative to uniform allocation, on
(i) the stability of the updated gain and of the gains queried around
it, and (ii) the convergence of $J_{\rm avg}$.
\end{problem}

We address the allocation in Section~\ref{sec:allocation} and the
stability and convergence parts in Section~\ref{sec:stability}.

\section{Personalized Rollout Allocation under a Communication Budget}
\label{sec:allocation}
To arrive at a variance-optimal allocation, we first solve the relaxation of Problem 1 over real counts,
\begin{equation}
 \min_{m\in\R^M}\ \sum\nolimits_{i=1}^M\frac{V_i}{m_i}
 \quad\text{s.t.}\quad m_i\ge m_{\min},\ \ \sum\nolimits_{i=1}^Mm_i\le S.
 \label{eq:relaxed-problem}
\end{equation}
 The
minimum count, which is imposed under the assumption $S\ge Mm_{\min}$, guarantees that every client contributes to each update
and, as shown in Section~\ref{sec:weights}, that its variance can be
estimated from the received coefficients, which requires
$m_{\min}\ge2$. Throughout this section the round index $k$ is fixed
and omitted from $V_{i,k}$ and $m_{i,k}$ when no confusion arises.

\begin{theorem}[Variance-optimal allocation]
\label{thm:allocation}
Let $S\ge Mm_{\min}$ and $V_i>0$ for all $i$. The optimal allocation problem~\eqref{eq:relaxed-problem}
has a unique solution $m^\star$. It exhausts the budget,
$\sum_im_i^\star=S$, and is given by
\begin{equation}
 m_i^\star=\max\big\{m_{\min},\sqrt{V_i/\lambda}\big\},
 \label{eq:optimal-counts}
\end{equation}
where $\lambda>0$ is chosen such that $\sum_im_i^\star=S$; this choice is
unique when $S>Mm_{\min}$. If no minimum count is active, then
\begin{equation}
 m_i^\star=\frac{S\sqrt{V_i}}{\sum_{j=1}^M\sqrt{V_j}},\quad
 \cV_k(m^\star)=\frac{\big(\sum_{i=1}^M\sqrt{V_i}\big)^2}{M^2S}
 \le\frac{\sum_{i=1}^MV_i}{MS},
 \label{eq:variance-optimum}
\end{equation}
where the right-hand side equals $\cV_k(m^{\rm eq})$ for the uniform
counts $m_i^{\rm eq}=S/M$, with equality if and only if all $V_i$
coincide.
\end{theorem}

The proof is given in Appendix~\ref{app:proofs}. The optimal counts
grow with the square root of the variance: a client whose estimates
have four times the variance of another receives twice as many rollout
pairs. At the optimum, every client with an inactive minimum count
offers the same marginal variance reduction per additional rollout
pair, which is the defining property of the Neyman allocation in
stratified sampling~\cite{neyman1934two}. We emphasize that the
allocation does not down-weight noisy clients; their weight in the
objective is fixed at $1/M$ by~\eqref{eq:global-ZO}, and the additional
rollout pairs serve to estimate their local means more accurately.

\begin{remark}[Static versus dynamic allocation]
\label{rem:dynamic}
The optimal counts in~\eqref{eq:optimal-counts} depend on the round
through the variances $V_{i,k}$. Since both the directional-estimation
variance and the rollout-noise variance depend on the current gain
$K_k$, the variances generally change as learning progresses.
Consequently, their relative values need not remain fixed, and the
rollout allocation should be recomputed at each round. A useful special
case arises when the clients differ only through the scale of their
process noise, $W_i=\alpha_iW_0$. Under the conditions of
Proposition~\ref{prop:scaled}, the variances then satisfy
$V_{i,k}=\alpha_i^2V_{0,k}$ for a common factor $V_{0,k}$, so the
ratios $V_{i,k}/V_{j,k}$ do not depend on $k$. Since the optimal
allocation depends on the variances only through their ratios, it is
then static; when no minimum count binds, it reduces to the
proportional rule $m_i^\star\propto\alpha_i$. In practice, however, the
server does not know the coefficients $\alpha_i$ and must estimate the
variances from the communicated coefficients, as described in
Section~\ref{sec:weights}. Recomputing the allocation each round then
allows these estimates to improve over time and tracks any changes in
operating conditions, without additional communication.
\end{remark}

\vspace{-0.05in}
\subsection{Integer Allocation}

The counts in~\eqref{eq:optimal-counts} are in general not integers.
To obtain an integer allocation, we observe that assigning one
additional rollout pair to client $i$ reduces the objective
of~\eqref{eq:relaxed-problem} by
\begin{equation}
 \rho_i(m_i)=\frac{V_i}{m_i}-\frac{V_i}{m_i+1}=\frac{V_i}{m_i(m_i+1)},
 \label{eq:marginal-gain}
\end{equation}
which is the measure of client effectiveness introduced in
Section~\ref{sec:intro}: it increases with the variance of the client's
estimates and decreases with the number of rollout pairs already
assigned to it. The routine $\mathsf{Alloc}$ invoked in
Algorithm~\ref{alg:fedlqr} uses this measure greedily: after reserving
$m_{\min}$ rollout pairs for every client, it solicits each remaining
pair from the currently most effective client, as summarized in
Algorithm~\ref{alg:alloc}.

\begin{algorithm}[t]
\small
\caption{$\mathsf{Alloc}(S,\{V_i\}_{i=1}^M,m_{\min})$: greedy integer
allocation}
\label{alg:alloc}
\begin{algorithmic}[1]
\Require budget $S\ge Mm_{\min}$; variances $V_i>0$; minimum count
$m_{\min}$.
\State $m_i\gets m_{\min}$ for all $i=1,\ldots,M$
\Comment{minimum participation}
\For{$n=1,\ldots,S-Mm_{\min}$}
  \State $i^\star\gets\arg\max_{i}\ \rho_i(m_i)$ by~\eqref{eq:marginal-gain}
  \Comment{most effective client}
  \State $m_{i^\star}\gets m_{i^\star}+1$
\EndFor
\State \Return $m=(m_1,\ldots,m_M)$
\end{algorithmic}
\end{algorithm}

\begin{lemma}[Optimality of the greedy rule $\mathsf{Alloc}$ in
Algorithm~\ref{alg:alloc}]
\label{lem:greedy}
For $S\ge Mm_{\min}$ and $V_i>0$, the allocation $m^{\rm opt}$ returned
by Algorithm~\ref{alg:alloc} minimizes $\sum_iV_i/m_i$ over all integer vectors
with $m_i\ge m_{\min}$ and $\sum_im_i\le S$.
\end{lemma}
\begin{proof}
Since every $\rho_i(\cdot)$ is positive, any optimal allocation
exhausts the budget, and the objective of an allocation $m$ equals
$\sum_iV_i/m_{\min}$ minus the sum of the marginal reductions
$\rho_i(m_{\min}),\ldots,\rho_i(m_i-1)$ over all $i$. Because
$\rho_i(\cdot)$ is strictly decreasing, the reductions available to
client $i$ must be collected in order, and the greedy rule selects the
$S-Mm_{\min}$ largest elements of the multiset
$\{\rho_i(n):i\le M,\ n\ge m_{\min}\}$ subject to this ordering, which
is the largest attainable total reduction.
\end{proof}

A direct implementation of greedy procedure in Algorithm~\ref{alg:alloc} requires $O(MS)$
operations per round; a priority queue reduces this to
$O(M+S\log M)$. Both are negligible compared with the cost of
executing the rollouts. To quantify the benefit of personalized
allocation relative to uniform counts, we define
\begin{equation}
 \Gamma_k=\frac{M\sum_{i=1}^MV_{i,k}}{\big(\sum_{i=1}^M\sqrt{V_{i,k}}\big)^2},
 \qquad
 \Gamma_k(S)=\frac{\cV_k(m^{\rm eq})}{\cV_k(m^{\rm opt})},
 \label{eq:stability-gain}
\end{equation}
where $S$ is assumed to be a multiple of $M$ so that $m^{\rm eq}$ is
integer. By~\eqref{eq:variance-optimum}, $\Gamma_k$ is the variance
ratio between uniform counts and the proportional allocation
$m_i\propto\sqrt{V_{i,k}}$, and $\Gamma_k(S)$ is the corresponding
ratio for the integer allocation actually used by the server. The
following properties relate the two.

\begin{proposition}[Variance-reduction factors]
\label{prop:gamma}
Let $V_{i,k}>0$ for all $i$ and $M\ge2$. Then
$1\le\Gamma_k(S)\le\Gamma_k<M$. Moreover, $\Gamma_k=1$ if and only if
all $V_{i,k}$ coincide, and $\Gamma_k(S)=\Gamma_k$ if and only if the
proportional counts $S\sqrt{V_{i,k}}/\sum_j\sqrt{V_{j,k}}$ are integers
no smaller than $m_{\min}$.
\end{proposition}
\begin{proof}
The uniform allocation is feasible for the integer problem, so
$\cV_k(m^{\rm opt})\le\cV_k(m^{\rm eq})$ and $\Gamma_k(S)\ge1$. The
integer optimum is feasible for the relaxation
of~\eqref{eq:relaxed-problem} without minimum counts, whose minimizer is
the proportional allocation; hence
$\cV_k(m^{\rm opt})\ge(\sum_i\sqrt{V_{i,k}})^2/(M^2S)$, which gives
$\Gamma_k(S)\le\Gamma_k$, with equality exactly when the proportional
allocation is itself feasible for the integer problem, since the
relaxed minimizer is unique. The bound $\Gamma_k\ge1$ and its equality
case follow from the Cauchy--Schwarz inequality
$(\sum_i\sqrt{V_{i,k}})^2\le M\sum_iV_{i,k}$, and $\Gamma_k<M$ follows
from $(\sum_i\sqrt{V_{i,k}})^2=\sum_iV_{i,k}+\sum_{i\ne j}
\sqrt{V_{i,k}V_{j,k}}>\sum_iV_{i,k}$.
\end{proof}

The factor $\Gamma_k$ depends only on the dispersion of the variances
across clients and approaches $M$ when a single client carries nearly
all of the uncertainty; the factor $\Gamma_k(S)$ additionally accounts
for the budget, the integrality of the counts, and the minimum
participation, and it is the quantity realized by
Algorithm~\ref{alg:fedlqr}. Section~\ref{sec:stability} shows that both
factors translate directly into stability and convergence guarantees.

\subsection{Estimating the Allocation Weights}
\label{sec:weights}

The allocation requires the variances $V_{i,k}$, which are unknown to
the server. They can, however, be estimated from the coefficients that
the server already receives. Conditional on $\cH_k$, the estimates
$Z_{i,k,1},\ldots,Z_{i,k,m_{i,k}}$ of client $i$ are independent and
identically distributed, so their sample variance is unbiased for
$V_{i,k}$. Since $\fro{U_{i,k,s}}=1$, it can be computed from the
scalars alone as
\begin{equation}
 \widehat V_{i,k}
 =\frac{D^2\sum_{s=1}^{m_{i,k}}d_{i,k,s}^2-m_{i,k}\fro{\widehat G_{i,k}}^2}
       {m_{i,k}-1},
 \qquad m_{i,k}\ge2,
 \label{eq:sample-variance}
\end{equation}
which satisfies $\E[\widehat V_{i,k}\mid\cH_k]=V_{i,k}$ and requires no
additional upload. The requirement $m_{i,k}\ge2$ is the reason for
imposing $m_{\min}\ge2$ in Algorithm~\ref{alg:fedlqr}. The server uses
$\widetilde V_{i,k+1}=\max\{v_{\rm floor},\widehat V_{i,k}\}$ as its
prediction of $V_{i,k+1}$, where the floor $v_{\rm floor}>0$ prevents a
client whose sample variance happens to be very small from being
assigned the minimum count indefinitely.

The prediction is subject to two sources of error: the sample variance
is computed from as few as $m_{\min}$ coefficients, and the gain changes
between rounds. The following result shows that the allocation degrades
gracefully with the relative accuracy of the predictions.

\begin{proposition}[Robustness to estimated weights]
\label{prop:robust}
Let $\widetilde V_i$ satisfy $(1-\epsilon)V_i\le\widetilde V_i\le
(1+\epsilon)V_i$ for all $i$ and some $0\le\epsilon<1$, and let
$\widehat m$ be the integer allocation computed by Algorithm~\ref{alg:alloc}
from $\{\widetilde V_i\}$. Then
\begin{equation}
 \cV_k(\widehat m)\le\frac{1+\epsilon}{1-\epsilon}\,\cV_k(m^{\rm opt}).
 \label{eq:allocation-robustness}
\end{equation}
\end{proposition}
\begin{proof}
Write $f(m)=\sum_iV_i/m_i$ and $\tilde f(m)=\sum_i\widetilde V_i/m_i$,
so that $(1-\epsilon)f\le\tilde f\le(1+\epsilon)f$ on all allocations.
By Lemma~\ref{lem:greedy}, $\widehat m$ minimizes $\tilde f$ over the
integer feasible set, which contains $m^{\rm opt}$. Hence
$f(\widehat m)\le\tilde f(\widehat m)/(1-\epsilon)
\le\tilde f(m^{\rm opt})/(1-\epsilon)
\le(1+\epsilon)f(m^{\rm opt})/(1-\epsilon)$, and the claim follows
from~\eqref{eq:centered-error}.
\end{proof}

Proposition~\ref{prop:robust} shows that predictions with a moderate
relative error preserve most of the variance reduction: for
$\epsilon=0.2$, the factor $(1+\epsilon)/(1-\epsilon)=1.5$ is small
compared with the values of $\Gamma_k(S)$ realized in
Section~\ref{sec:numerics}. We note, however, that unbiasedness
of~\eqref{eq:sample-variance} does not by itself guarantee a small
$\epsilon$, since the estimate is formed from few samples and the
variance may change with the gain. The guarantees of
Section~\ref{sec:stability} are therefore stated for the true
variances, and Section~\ref{sec:numerics} evaluates the allocation with
estimated weights directly.

\section{Stability and Convergence with Personalized Rollouts}
\label{sec:stability}

This section addresses the stability and convergence parts of
Problem~\ref{prob:allocation}. Process noise does not affect whether a
fixed gain is stabilizing, which depends on $A-BK$ alone; it affects
the random update. A single update can leave the stabilizing set even
when its conditional mean would remain inside, and the sampling error
that survives averaging limits how closely repeated updates approach
the optimum. Both effects are governed by the variance
$\cV_k(m)$ of Section~\ref{sec:allocation}, and both therefore improve
by the variance-reduction factor of the allocation. All statements
below condition on the history $\cH_k$, which fixes $K_k$, $m_k$, and
$\eta_k$ before the coefficients of round $k$ are collected. Proofs are
given in Appendix~\ref{app:proofs}.

\subsection{Stability of a Single Update}

Suppose that the closed ball of radius $d_k>r$ about $K_k$ is contained
in $\cS$; such stabilizing neighborhoods are standard in LQR policy
optimization~\cite{fazel2018global,song2025stochastic}. Since the
perturbed gains $K_{k+1}\pm rU$ queried in the next round must also be
stabilizing, the update may move the gain by at most $d_k-r$. The
conditional mean of the step has length $\eta_k\fro{\overline G_k}$, so
the sampling error $\Xi_k=\widehat G_k-\overline G_k$ may consume at
most the tolerance
\begin{equation}
 \tau_k=\frac{d_k-r}{\eta_k}-\fro{\overline G_k}.
 \label{eq:stability-tolerance}
\end{equation}
If $\tau_k>0$ and $\fro{\Xi_k}\le\tau_k$, then $K_{k+1}$ and all gains
within distance $r$ of it are stabilizing. The rollout counts affect the
distribution of $\Xi_k$ but neither $\overline G_k$ nor $\tau_k$, which
is what allows allocations to be compared on equal terms.

\begin{theorem}[One-round stability]
\label{thm:stability}
Let $\tau_k>0$, and let $\mathsf F_k$ denote the event that $K_{k+1}$
or some gain $K_{k+1}+E$ with $\fro E\le r$ is not stabilizing. Then
\begin{equation}
 \Pr(\mathsf F_k\mid\cH_k)\le\frac{\cV_k(m)}{\tau_k^2}.
 \label{eq:failure-certificate}
\end{equation}
In particular, $\cV_k(m)\le\delta_k\tau_k^2$ certifies, with
confidence $1-\delta_k$, that the next gain and all gains queried
around it are stabilizing.
\end{theorem}

The bound relies on second moments only, so it requires no
distributional assumption on the coefficients beyond
Assumption~\ref{ass:noise}, at the price of being conservative. If
$\tau_k\le0$, no reduction of variance can certify the update and the
step size must be reduced. In practice, an upper bound on
$\fro{\overline G_k}$, for instance the bound $G$ introduced in
Section~\ref{sec:convergence}, may replace the exact norm
in~\eqref{eq:stability-tolerance}.

Because neither $\overline G_k$ nor $\tau_k$ depends on the counts,
Theorem~\ref{thm:stability} compares personalized and uniform counts at
the same gain, radius, and rollout procedure. The comparison can be
expressed in terms of the failure bound, the sufficient budget, or the
certified step size.

\begin{corollary}[Stability gain]
\label{cor:stability-gain}
Let $\tau_k>0$ and let $S$ be a multiple of $M$. Relative to the
uniform counts $m^{\rm eq}$, the integer allocation $m^{\rm opt}$
of Algorithm~\ref{alg:alloc} satisfies the following.
\begin{enumerate}
\item[(i)] At the same budget $S$, the failure bound
\eqref{eq:failure-certificate} is divided by $\Gamma_k(S)$.
\item[(ii)] At the same confidence $1-\delta_k$, if no minimum count is
active and the proportional counts are integers, the budgets
\begin{equation}
 S_k^{\rm per}=\frac{(\sum_i\sqrt{V_{i,k}})^2}{M^2\delta_k\tau_k^2},
 \qquad S_k^{\rm eq}=\Gamma_kS_k^{\rm per},
 \label{eq:sufficient-budgets}
\end{equation}
are sufficient for the certificate under personalized and uniform
counts, respectively; for $S_k^{\rm per}\le S<S_k^{\rm eq}$ the
certificate holds for personalized counts only.
\item[(iii)] At the same budget and confidence, the largest certified
step size is
\begin{subequations}\label{eq:certified-step}
\begin{align}
 &\eta_k^{\max}(m)=\frac{d_k-r}
 {\fro{\overline G_k}+\sqrt{\cV_k(m)/\delta_k}},\\
 &1\le\frac{\eta_k^{\max}(m^{\rm opt})}{\eta_k^{\max}(m^{\rm eq})}
 \le\sqrt{\Gamma_k(S)},
\end{align}
\end{subequations}
so that every step size between the two values is certified for
personalized counts only.
\end{enumerate}
\end{corollary}

Item (iii) is the stability margin that the experiments of
Section~\ref{sec:numerics} test directly. When no minimum count binds
and the proportional counts are integers, $\Gamma_k(S)$ coincides with
$\Gamma_k$ by Proposition~\ref{prop:gamma}; in general $\Gamma_k(S)$ is
the factor realized by the server.

The sufficient budget in~\eqref{eq:sufficient-budgets} assumes exact
variances and ignores the minimum participation. The following
corollary accounts for both.

\begin{corollary}[Sufficient budget with minimum counts and estimated
weights]
\label{cor:budget}
Let the predicted variances satisfy the relative-error condition of
Proposition~\ref{prop:robust} with $0\le\epsilon<1$, and let
$m_{\min}\ge1$. Then the budget
\begin{equation}
 S\ge Mm_{\min}+\left\lceil
 \frac{(1+\epsilon)(\sum_i\sqrt{V_{i,k}})^2}
 {(1-\epsilon)M^2\delta_k\tau_k^2}\right\rceil
 \label{eq:integer-stability-budget}
\end{equation}
is sufficient for the certificate of Theorem~\ref{thm:stability} under
the integer allocation computed from the predicted variances.
Moreover, this allocation improves on uniform counts whenever
$\Gamma_k(S)>(1+\epsilon)/(1-\epsilon)$.
\end{corollary}

The first term in~\eqref{eq:integer-stability-budget} reserves the
minimum participation of every client; the second is the continuous
budget $S_k^{\rm per}$ inflated by the prediction error, and it reduces
to $S_k^{\rm per}$ for exact weights.

\subsection{The Case of Scaled Process Noise}

The factor $\Gamma_k$ admits a closed form when the clients' noise
processes differ only in scale, which is also the setting of the
experiments in Section~\ref{sec:numerics}.

\begin{proposition}[Scaled process noise]
\label{prop:scaled}
Let every rollout start from $x_0=0$, and let
$w_t^{(i)}\overset{d}{=}\sqrt{\alpha_i}\,w_t^{(0)}$ for a reference
noise process $w_t^{(0)}$ and scales $\alpha_i>0$, so that
$W_i=\alpha_iW_0$. Then, for every gain and every round,
$\overline G_{i,k}=\alpha_i\overline G_{0,k}$,
$V_{i,k}=\alpha_i^2V_{0,k}$, the proportional counts are
$m_{i,k}^\star=S\alpha_i/\sum_j\alpha_j$, and, with
$\bar\alpha=\frac{1}{M
}\sum_i\alpha_i$,
\begin{equation}
 \Gamma_k=1+\frac{\frac{1}{M}\sum_{i=1}^M(\alpha_i-\bar\alpha)^2}{\bar\alpha^2}.
 \label{eq:noise-heterogeneity-gain}
\end{equation}
Furthermore, under the proportional counts, when they are integers no
smaller than $m_{\min}$, the aggregated update $\widehat G_k$ has the
same conditional distribution as the update of a homogeneous fleet of
$M$ clients with noise covariance $\bar\alpha W_0$ and uniform counts
$S/M$.
\end{proposition}

The gain is therefore one plus the squared coefficient of variation of
the noise scales: it is determined by their dispersion, not by their
level, and it is independent of the gain and of the round, so that
$\Gamma_k(S)$ is likewise constant along the learning trajectory. The
mechanism is transparent. By linearity, the trajectory of client $i$
is $\sqrt{\alpha_i}$ times a reference trajectory, so each of its
coefficients is $\alpha_i$ times a reference coefficient. Every
allocation thus averages the same $S$ reference coefficients, with the
copies of client $i$ weighted by $\alpha_i/(Mm_{i,k})$, and the sum of
the weights equals $\bar\alpha$ for every allocation. Proportional
counts are the only allocation with uniform weights, which minimizes
the variance of a weighted sum with fixed total weight; uniform counts
weight the noisiest clients most heavily and inflate the variance by
$\Gamma_k$. When the proportional counts are not feasible, because of
integrality or minimum participation, the integer allocation realizes
the smaller factor $\Gamma_k(S)$ and the equivalence with a homogeneous
fleet holds only approximately. We also note that raising all scales
together increases $\fro{\overline G_k}$ and hence reduces $\tau_k$,
so that a budget threshold is meaningful only for a specified step size
and margin.
\vspace{-0.05in}
\begin{remark}[Beyond the second-moment bound]
\label{rem:gaussian}
The bound~\eqref{eq:failure-certificate} controls the escape
probability through $\cV_k(m)$ alone. Under scaled noise, the sampling
error is a weighted sum $\Xi_k\overset{d}{=}\sum_{j=1}^Sw_j\widetilde
Z_j$ of $S$ independent and identically distributed centered reference
estimates, with $\|w\|_2^2=\cV_k(m)/V_{0,k}$. The estimates
$\widetilde Z_j$ are not Gaussian, since each is a scalar multiple of a
direction on the unit sphere, but for large $S$ the central limit
theorem suggests that $\Xi_k$ is approximately Gaussian with covariance
proportional to $\|w\|_2^2$. Under this approximation the escape
probability $\Pr(\fro{\Xi_k}>\tau_k\mid\cH_k)$ is strictly increasing
in $\cV_k(m)$, so that proportional counts minimize the probability
itself and not only its bound. The experiments of
Section~\ref{sec:numerics} examine the accuracy of this approximation.
\end{remark}
\vspace{-0.1in}\vspace{-0.1in}\vspace{-0.1in}
\subsection{Convergence to an Accuracy Neighborhood}
\label{sec:convergence}

We now turn to the behavior of the learning process over many rounds.
Write $J=J_{\rm avg}$, $\bar W=M^{-1}\sum_iW_i$, and
$J^\star=\min_{\cS}J$, and assume $\bar W\succ0$; individual $W_i$ may
be singular. Since $J(K)=M^{-1}\sum_i\tr(P_KW_i)=\tr(P_K\bar W)$, the
fleet objective is itself an LQR cost with noise covariance $\bar W$.
It is minimized by the Riccati gain $K^\star$, which minimizes every
$J_i$ simultaneously because $P_{K^\star}\preceq P_K$ for all
$K\in\cS$, and it satisfies the gradient-domination inequality of
LQR on all of $\cS$~\cite[Lemma~3.12]{bu2019lqr},
\begin{equation}
 \fro{\nabla J(K)}^2\ge2\mu_0\bigl(J(K)-J^\star\bigr),\quad
 \mu_0=\frac{2\lambda_{\min}(R)\lambda_{\min}(\bar W)^2}{\|\Sigma_\star\|_2},
 \label{eq:aggregate-pl}
\end{equation}
where $\Sigma_\star$ solves
$\Sigma_\star=\bar W+A_{K^\star}\Sigma_\star A_{K^\star}^\top$. Two
features of~\eqref{eq:aggregate-pl} are specific to the present
setting. First, no heterogeneity term appears: unlike federated LQR
with heterogeneous dynamics~\cite{wang2023federated}, all clients share
the minimizer $K^\star$, so the fleet gradient vanishes there. Second,
the constant $\mu_0$ does not degrade near the boundary of $\cS$,
because the state covariance $\Sigma_K$ satisfies $\Sigma_K\succeq\bar W$
for every stabilizing $K$. Only the smoothness constant and the
stability margin below are local.

Let $b_k=\overline G_k-\nabla J(K_k)$ denote the oracle bias, which
collects the smoothing bias of the finite radius and any bias of the
cost measurement, and which the counts do not affect. Fix $c>J^\star$
with $J(K_0)\le c$ and consider the sublevel set
$\mathcal C=\{K\in\cS:J(K)\le c\}$, which is
compact~\cite{fazel2018global,bu2019lqr}. Choose $d>r$ such that the
closed ball of radius $d$ about every $K\in\mathcal C$ lies in $\cS$,
let $L$ bound the operator norm of the Hessian of $J$ on the union of
these balls, let $G$ bound $\fro{\nabla J}$ on $\mathcal C$, and fix
$0<\mu\le\min\{\mu_0,L\}$.

\begin{theorem}[Stable convergence]
\label{thm:convergence}
Let the step size $\eta$ be constant, and suppose that on every
history whose iterates lie in $\mathcal C$, the bias and variance
satisfy $\fro{b_k}\le\bar b$ and $\cV_k(m_k)\le\bar v$ for $k<N$. For
$0<\delta<1$, define $s_N=\sqrt{N\bar v/\delta}$ and
$R_N=(\bar b+s_N)^2/(2\mu)$. If
\begin{equation}
 0<\eta\le\min\Big\{\frac1L,\ \frac{d-r}{G+\bar b+s_N}\Big\}
 \quad\text{and}\quad R_N\le c-J^\star,
 \label{eq:convergence-conditions}
\end{equation}
then, with probability at least $1-\delta$, all $N$ updated gains and
all gains queried around them are stabilizing, all iterates remain in
$\mathcal C$, and
\begin{equation}
 J(K_k)-J^\star\le(1-\mu\eta)^k\bigl(J(K_0)-J^\star\bigr)+R_N,
 \qquad 0\le k\le N.
 \label{eq:linear-neighborhood}
\end{equation}
\end{theorem}

The error contracts geometrically at a rate determined by $\mu$ and
$\eta$ and settles in a neighborhood of radius $R_N$, which consists of
a bias floor $\bar b^2/(2\mu)$, set by the radius and the rollout
procedure, and a sampling term governed by $s_N$. The factor $N/\delta$
in $s_N$ arises because every round is controlled through second
moments and a union bound; as a consequence, the guarantee is a
finite-horizon one, in which the admissible step size and the
neighborhood depend on the number of rounds. Sharper concentration
would replace $N/\delta$ by a logarithmic factor, which we leave to
future work. The allocation enters Theorem~\ref{thm:convergence} only
through $\bar v$.
\vspace{-0.05in}
\begin{corollary}[Convergence gain]
\label{cor:convergence-gain}
Suppose that uniform counts admit the variance bound $\bar v_{\rm eq}$
in Theorem~\ref{thm:convergence}, and that $\Gamma_k(S)\ge\Gamma$ for
all $k<N$ and all histories in $\mathcal C$, where $\Gamma\ge1$; under
scaled noise with feasible proportional counts, $\Gamma$ may be taken
as~\eqref{eq:noise-heterogeneity-gain}. Then personalized counts admit
$\bar v_{\rm per}=\bar v_{\rm eq}/\Gamma$ and
$s_{N,\rm per}=s_{N,\rm eq}/\sqrt{\Gamma}$, and the following hold.
\begin{enumerate}
\item[(i)] At the same step size, the neighborhood shrinks from
$(\bar b+s_{N,\rm eq})^2/(2\mu)$ to
$(\bar b+s_{N,\rm eq}/\sqrt\Gamma)^2/(2\mu)$, that is, by the factor
$\Gamma$ when the bias is negligible.
\item[(ii)] At the same budget, the step size admitted
by~\eqref{eq:convergence-conditions} grows by a factor in
$[1,\sqrt\Gamma]$, and with it the certified contraction per round.
\item[(iii)] Let $\varepsilon\le2(c-J^\star)$, let
$N\ge\log\big(2(J(K_0)-J^\star)/\varepsilon\big)/(\mu\eta)$, and
suppose $\cV_k(m_k)\le a/S$ on $\mathcal C$. If
$t_\varepsilon=\min\{\sqrt{\mu\varepsilon}-\bar b,\,(d-r)/\eta-G-\bar b\}>0$,
then a per-round budget $S\ge Na/(\delta t_\varepsilon^2)$ yields
$J(K_N)-J^\star\le\varepsilon$ with probability at least $1-\delta$,
and the constant $a$ is $\Gamma$ times smaller for personalized than
for uniform counts.
\end{enumerate}
\end{corollary}

Item (iii) shows that the communication required to reach a prescribed
accuracy is divided by the same factor that governs one-round
stability. With estimated weights, Proposition~\ref{prop:robust}
replaces $\Gamma$ by $\Gamma(1-\epsilon)/(1+\epsilon)$ throughout. The
bias floor is unchanged by the allocation; reducing it requires a
smaller radius or a longer rollout, both of which increase the variance
and hence the value of allocating rollouts where they are most
effective.

\vspace{-0.1in}\vspace{-0.1in}
\section{Numerical Results}
\label{sec:numerics}

This section evaluates the predictions of
Corollaries~\ref{cor:stability-gain} and~\ref{cor:convergence-gain} and
of Proposition~\ref{prop:scaled} on simulated trajectory rollouts. We
consider fleets of ten and twenty clients with common dynamics and
costs
\begin{equation}
 A=\begin{bmatrix}
 1.08&0.03&0&0\\0.03&1.12&0.03&0\\
 0&0.03&1.16&0.03\\0&0&0.03&1.20
 \end{bmatrix},\quad
 \begin{aligned}B&=I_4,\\Q&=1.5I_4,\\R&=0.4I_4,\end{aligned}
 \label{eq:numerical-system}
\end{equation}
so that $D=16$. Every rollout starts from $x_0=0$ and is driven by
Gaussian process noise with covariance $\alpha_iI_4$, where
$\alpha_i=1-h$ for the first $90\%$ of the clients and
$\alpha_i=1+9h$ for the remaining $10\%$. The mean covariance equals
$I_4$ for every heterogeneity level $h\in[0,1)$, so that $h$ changes
the reliability of the measurements without changing the fleet
objective, its optimal value $J_{\rm avg}(K^\star)=7.72769$, or the
initial gap $J_{\rm avg}(K_0)-J_{\rm avg}(K^\star)=1.23916$ at
$K_0=1.30I_4$. Unless stated otherwise, $h=0.8$, for which the noise
scales are $0.2$ and $8.2$, the factor~\eqref{eq:noise-heterogeneity-gain}
is $\Gamma_k=1+9h^2=6.76$, and the integer factor at $S=4M$ with
$m_{\min}=2$ is $\Gamma_k(S)=5.22$; the noisiest client is then assigned
$22$ of the $40$ rollout pairs instead of four.

All methods use the radius $r=0.18$ and $m_{\min}=2$. Each cost
measurement averages $160$ stage costs, and the two rollouts of a pair
use independent noise realizations. We compare uniform counts
(\emph{Uniform}) with Algorithm~\ref{alg:fedlqr} (\emph{Personalized}),
whose weights are estimated from the coefficients of the preceding
round after a calibration batch of four coefficients per client at
$K_0$; this batch equals the first-round budget $S/M$ at $S=4M$ and is
charged to the communication total in the learning experiments. A
reference implementation that uses the exact weights $\alpha_i^2$ was
indistinguishable from the estimated weights in every experiment
reported below and is omitted. Random number streams are paired across
methods, no update is projected, clipped, or replaced, and the system
matrices are used offline only to evaluate costs and to classify gains
as stabilizing or not.
\vspace{-0.1in}\vspace{-0.1in}
\subsection{Variance Reduction}

We first verify the variance-reduction factor. With $3{,}000$
independent calibration-and-evaluation trials at $K_0$ and $S=4M$, the
measured ratio $\cV_k(m^{\rm eq})/\cV_k(\widehat m)$ of the sampling-error
variances is $1.23$, $2.26$, $3.92$, $5.26$, and $5.37$ at
$h=0.2,0.4,0.6,0.8,0.9$ for ten clients, and $1.22$, $2.22$, $3.93$,
$4.69$, and $4.95$ for twenty, against the predicted
$\Gamma_k(S)=1.35$, $2.39$, $4.11$, $5.22$, and $5.44$. The prediction
follows the closed form $1+9h^2$ only while no minimum count binds.
From $h\approx0.6$ onward, the noisiest client already receives the
$22$ rollout pairs that remain after the reserve of two per client, so
the realized factor saturates near $5.4$ although $1+9h^2$ reaches
$8.3$; this is the gap between $\Gamma_k$ and $\Gamma_k(S)$ described
in Proposition~\ref{prop:gamma}. The measured ratios for twenty clients
fall about $10\%$ below the prediction at $h\ge0.8$, which we attribute
to the estimation error of weights formed from two coefficients. At
$h=0$, where uniform counts are optimal, the estimated weights cost
$14\%$ and $12\%$ of variance for ten and twenty clients, respectively,
for the same reason.

\begin{figure}[t]
 \centering
 \includegraphics[scale=0.75]{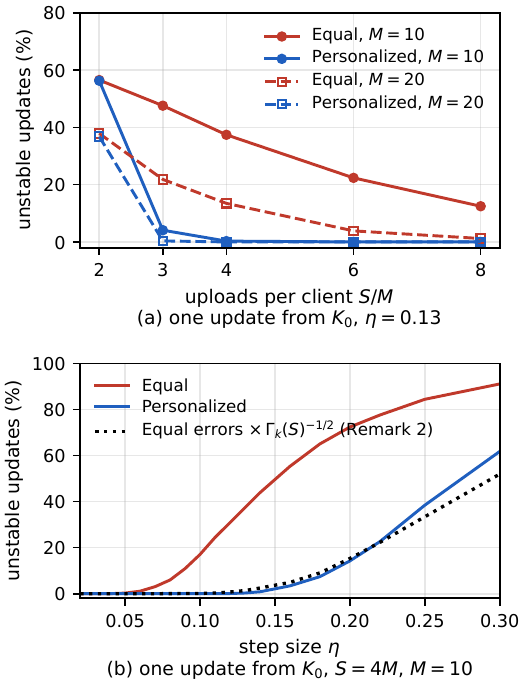}
\caption{Percentage of destabilizing updates after one step from $K_0$
at $h=0.8$, under uniform counts (``Equal'') and personalized counts
(Algorithm~\ref{alg:fedlqr}). (a) Versus the number of rollout pairs per
client, $S/M$, at $\eta=0.13$, for $M=10$ and $M=20$. (b) Versus the
step size at $S=4M$ and $M=10$. The dotted curve in (b) is obtained by
scaling the uniform-count sampling errors by $\Gamma_k(S)^{-1/2}$ and
recomputing the failures (Remark~\ref{rem:gaussian}).}
 \label{fig:stability}
\end{figure}

\begin{figure}[t]
 \centering
 \includegraphics[scale=0.75]{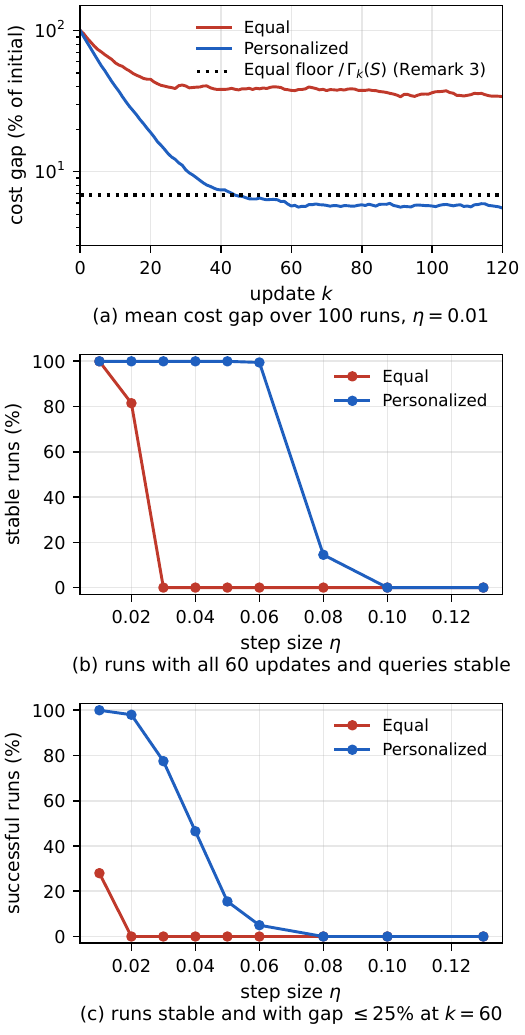}
\caption{Learning with ten clients at $h=0.8$ and $S=4M$, under uniform
counts (``Equal'') and personalized counts (Algorithm~\ref{alg:fedlqr}).
(a) Mean cost gap over $100$ runs at $\eta=0.01$; the dashed and dotted
curves use the larger step sizes $\eta=0.02$ and $0.03$, which only
personalized counts tolerate, and the horizontal line marks the
uniform-count floor divided by $\Gamma_k(S)$
(Corollary~\ref{cor:convergence-gain}; labeled Remark~3 in the legend).
(b) Fraction of $200$ runs in which all $60$ updates and all queried
gains are stabilizing, versus step size. (c) Fraction of runs that are
stable and within $25\%$ of the initial gap after $60$ updates, versus
step size.}
 \label{fig:learning}
\end{figure}

\subsection{Stability of a Single Update}

Fig.~\ref{fig:stability}(a) applies one update from $K_0$ with
$\eta=0.13$ and reports the percentage of $3{,}000$ trials in which
$\rho(A-BK_1)\ge1$. For ten clients, uniform counts produce a
destabilizing gain in $56.5$, $47.6$, $37.4$, $22.4$, and $12.5\%$ of
the trials at $S/M=2,3,4,6,8$, and personalized counts in $56.3$,
$4.1$, $0.3$, $0$, and $0\%$. For twenty clients, the corresponding
figures are $37.9$, $21.8$, $13.5$, $3.9$, and $1.2\%$ against
$36.8$, $0.4$, $0$, $0$, and $0\%$. At $S=2M$ the minimum counts force
identical allocations, and the small difference between the two
methods reflects the calibration batch, which precedes the measured
update only for the personalized method; the separation appears as
soon as the budget can be redistributed. The conditional mean of the
update is the same for both methods, and its spectral radius lies
between $0.10$ and $0.11$ at $S=4M$, so the failures are caused by
sampling error alone.

Fig.~\ref{fig:stability}(b) fixes $S=4M$ and sweeps the step size,
which corresponds to the certified-step comparison
in~\eqref{eq:certified-step}. Uniform counts begin to fail at
$\eta\approx0.06$ and reach a $37\%$ failure rate at $\eta=0.13$,
whereas personalized counts begin to fail at $\eta\approx0.14$ and
reach $37\%$ only at $\eta=0.25$. The step sizes at which the failure
rate reaches $1\%$, $5\%$, and $10\%$ are larger by the factors
$2.40$, $2.19$, and $2.13$, close to $\sqrt{\Gamma_k(S)}=2.29$. The
dotted curve tests the approximation of Remark~\ref{rem:gaussian}: each
uniform-count sampling error is multiplied by $\Gamma_k(S)^{-1/2}$,
which is the law of the personalized error under scaled noise and
proportional counts, and the failure rate is recomputed. It tracks the
personalized curve within two percentage points up to $\eta=0.2$ and is
slightly pessimistic, because with only four coefficients from the
noisiest client the uniform-count error has a heavier tail than its
variance alone suggests. We emphasize that these experiments measure
the actual instability rate, not the sufficient thresholds of
Theorem~\ref{thm:stability}; for $0$ failures in $3{,}000$ trials, the
one-sided $95\%$ upper confidence bound on the failure probability is
approximately $0.1\%$.

\subsection{Learning over Multiple Rounds}

Fig.~\ref{fig:learning} reports learning runs with ten clients at
$h=0.8$ and $S=4M$. Fig.~\ref{fig:learning}(a) shows the mean cost gap
over $100$ runs of $120$ updates at $\eta=0.01$, a step size at which
every run of both methods remains stable. The two methods contract at
the same rate during the first $40$ updates, as the common conditional
mean of the update implies, and then settle at different floors:
$35.6\%$ of the initial gap for uniform counts and $5.8\%$ for
personalized counts. The observed reduction of $84\%$ is consistent
with the value $1-1/\Gamma_k(S)=81\%$ obtained by dividing the
sampling-error floor by $\Gamma_k(S)$, as suggested by
Corollary~\ref{cor:convergence-gain}(i); we regard this as a heuristic
check, since the corollary bounds a high-probability neighborhood
rather than the mean floor. The same runs at other heterogeneity levels
give uniform-count floors of $4.4$, $5.8$, $11.0$, $20.6$, $35.6$, and
$49.0\%$ at $h=0,0.2,0.4,0.6,0.8,0.9$, whereas the personalized floors
remain between $4.5\%$ and $6.6\%$. Personalized allocation thus
removes most of the effect of noise heterogeneity on the learning
floor, and the fleet learns nearly as well as a homogeneous fleet at
the mean noise level; the residual gap at large $h$ reflects the
minimum participation constraint, which prevents the proportional
allocation from being realized. The dashed and dotted curves in
Fig.~\ref{fig:learning}(a) use the larger step sizes $\eta=0.02$ and
$0.03$, which only personalized counts can afford: $90\%$ of the runs
reach the target gap of $25\%$ after $14$ updates instead of $21$ at
$\eta=0.01$, with floors of $12\%$ and $19\%$, respectively, whereas
uniform counts cannot use these step sizes at all.

Figs.~\ref{fig:learning}(b) and~(c) use $200$ runs of $60$ updates for
each step size. Fig.~\ref{fig:learning}(b) shows the fraction of runs
in which every update and every queried gain is stabilizing. Uniform
counts complete all $60$ updates in every run at $\eta=0.01$, in
$81.5\%$ of the runs at $\eta=0.02$, and in no run for $\eta\ge0.03$,
where $48$ runs first fail at a queried gain and $152$ at an update.
Personalized counts complete every run up to $\eta=0.05$, $99.5\%$ of
the runs at $\eta=0.06$, and $14.5\%$ at $\eta=0.08$. The usable step
size is therefore about five times larger, which exceeds the factor
$\sqrt{\Gamma_k(S)}=2.3$ certified in~\eqref{eq:certified-step}; the
certificate is only sufficient, and the uniform-count error has a
heavier tail than its variance indicates. Fig.~\ref{fig:learning}(c)
verifies that the stable runs also learn: a run is counted as
successful if it remains stable and its gap after $60$ updates is at
most $25\%$ of the initial gap. At $\eta=0.01$, the only step size at
which uniform counts are reliably stable, their floor exceeds the
target and $28\%$ of the runs succeed; at larger step sizes none do.
Personalized counts succeed in $100$, $98$, $78$, $47$, and $16\%$ of
the runs at $\eta=0.01,\dots,0.05$, where the decline reflects a floor
that grows with the step size, as in stochastic gradient methods.
Personalized allocation therefore attains, within a few hundred
transmitted scalars and over a range of step sizes, an accuracy target
that uniform allocation cannot reach at the same budget.
\section{Conclusion}
\label{sec:conclusion}

We have studied federated model-free learning of a common LQR gain by
clients that share dynamics and costs but experience process noise of
different intensities, under a per-round budget on transmitted scalars
that arises from bandwidth and energy constraints. Because a two-point
zeroth-order estimate separates into a data-dependent scalar and a
seed-reproducible direction, the budget on scalars is a budget on
trajectory rollouts, and its allocation across clients determines the
accuracy of every policy update. Allocating rollouts in proportion to
the standard deviation of each client's coefficients minimizes the
variance of the aggregated update, is implemented exactly by a greedy
integer rule that solicits each rollout pair from the currently most
effective client, and requires only variances that the server
estimates from the coefficients it already receives.

The benefit of the allocation is captured by a single factor
$\Gamma_k(S)$, which divides the variance of the update, the one-round
failure bound, the budget sufficient for a stability certificate, and
the sampling-error floor of the convergence guarantee, and whose square
root bounds the increase of the certified step size. Under scaled
process noise the factor equals one plus the squared coefficient of
variation of the noise scales, and when the proportional allocation is
feasible the fleet learns exactly as a homogeneous fleet at the mean
noise level would. The experiments reproduce these factors: the step
sizes at which instability appears grow by a factor between $2.1$ and
$2.4$ against the predicted $\sqrt{\Gamma_k(S)}=2.3$, and at a fixed
budget uniform allocation loses stability at step sizes that
personalized allocation tolerates.

Several extensions are left for future work. The guarantees are stated
for the true variances, whereas the server uses estimates; the
experiments indicate that estimated weights recover most of the gain,
but a propagation analysis of the estimation error through the
allocation and the policy iteration remains open. The certificates rely
on second moments, and sharper concentration would remove the factor
$N/\delta$ from the convergence neighborhood. Finally, a per-client cap
on rollouts, motivated by the exploration burden and energy expenditure
of individual systems, and fleets with heterogeneous dynamics are
natural extensions of the allocation problem.

\bibliographystyle{ieeetr}%
\bibliography{bib/alias,bib/Reference} %

@article{dean2020sample,
  author  = {Dean, Sarah and Mania, Horia and Matni, Nikolai and Recht, Benjamin and Tu, Stephen},
  title   = {On the sample complexity of the linear quadratic regulator},
  journal = {Foundations of Computational Mathematics},
  volume  = {20},
  number  = {4},
  pages   = {633--679},
  year    = {2020}
}

@inproceedings{fazel2018global,
  author    = {Fazel, Maryam and Ge, Rong and Kakade, Sham M. and Mesbahi, Mehran},
  title     = {Global convergence of policy gradient methods for the linear quadratic regulator},
  booktitle = {Proceedings of the 35th International Conference on Machine Learning},
  series    = {PMLR},
  volume    = {80},
  pages     = {1467--1476},
  year      = {2018}
}

@article{bu2019lqr,
  author  = {Bu, Jingjing and Mesbahi, Afshin and Fazel, Maryam and Mesbahi, Mehran},
  title   = {{LQR} through the lens of first order methods: Discrete-time case},
  journal = {arXiv preprint arXiv:1907.08921},
  year    = {2019}
}

@inproceedings{tu2019gap,
  author    = {Tu, Stephen and Recht, Benjamin},
  title     = {The gap between model-based and model-free methods on the linear quadratic regulator: An asymptotic viewpoint},
  booktitle = {Proceedings of the Conference on Learning Theory (COLT)},
  series    = {PMLR},
  volume    = {99},
  pages     = {3036--3083},
  year      = {2019}
}

@inproceedings{malik2019derivative,
  author    = {Malik, Dhruv and Pananjady, Ashwin and Bhatia, Kush and Khamaru, Koulik and Bartlett, Peter L. and Wainwright, Martin J.},
  title     = {Derivative-free methods for policy optimization: Guarantees for linear quadratic systems},
  booktitle = {Proceedings of the 22nd International Conference on Artificial Intelligence and Statistics (AISTATS)},
  series    = {PMLR},
  volume    = {89},
  pages     = {2916--2925},
  year      = {2019}
}

@article{mohammadi2022convergence,
  author  = {Mohammadi, Hesameddin and Zare, Armin and Soltanolkotabi, Mahdi and Jovanovi{\'c}, Mihailo R.},
  title   = {Convergence and sample complexity of gradient methods for the model-free linear-quadratic regulator problem},
  journal = {IEEE Transactions on Automatic Control},
  volume  = {67},
  number  = {5},
  pages   = {2435--2450},
  year    = {2022}
}

@article{song2025stochastic,
  author  = {Song, Bowen and Iannelli, Andrea},
  title   = {Convergence guarantees of model-free policy gradient methods for {LQR} with stochastic data},
  journal = {arXiv preprint arXiv:2502.19977},
  year    = {2025},
  note    = {revised Apr.~2026}
}

@article{duchi2015optimal,
  author  = {Duchi, John C. and Jordan, Michael I. and Wainwright, Martin J. and Wibisono, Andre},
  title   = {Optimal rates for zero-order convex optimization: The power of two function evaluations},
  journal = {IEEE Transactions on Information Theory},
  volume  = {61},
  number  = {5},
  pages   = {2788--2806},
  year    = {2015}
}

@article{shamir2017optimal,
  author  = {Shamir, Ohad},
  title   = {An optimal algorithm for bandit and zero-order convex optimization with two-point feedback},
  journal = {Journal of Machine Learning Research},
  volume  = {18},
  number  = {52},
  pages   = {1--11},
  year    = {2017}
}

@article{wang2023federated,
  author  = {Wang, Han and Toso, Leonardo F. and Mitra, Aritra and Anderson, James},
  title   = {Model-free learning with heterogeneous dynamical systems: A federated {LQR} approach},
  journal = {Transactions on Machine Learning Research},
  year    = {2026}
}

@inproceedings{toso2024meta,
  author    = {Toso, Leonardo F. and Zhan, Donglin and Anderson, James and Wang, Han},
  title     = {Meta-learning linear quadratic regulators: A policy gradient {MAML} approach for model-free {LQR}},
  booktitle = {Proceedings of the 6th Annual Learning for Dynamics and Control Conference (L4DC)},
  series    = {PMLR},
  volume    = {242},
  pages     = {902--915},
  year      = {2024}
}

@article{rostami2026scalar,
  author  = {Rostami, Mohammadreza and Talebi, Shahriar and Kia, Solmaz S.},
  title   = {Scalar federated learning for linear quadratic regulator},
  journal = {arXiv preprint arXiv:2604.05088},
  year    = {2026}
}

@article{recht2019tour,
  author  = {Recht, Benjamin},
  title   = {A tour of reinforcement learning: The view from continuous control},
  journal = {Annual Review of Control, Robotics, and Autonomous Systems},
  volume  = {2},
  pages   = {253--279},
  year    = {2019}
}

@article{hu2023toward,
  author  = {Hu, Bin and Zhang, Kaiqing and Li, Na and Mesbahi, Mehran and Fazel, Maryam and Ba{\c{s}}ar, Tamer},
  title   = {Toward a theoretical foundation of policy optimization for learning control policies},
  journal = {Annual Review of Control, Robotics, and Autonomous Systems},
  volume  = {6},
  pages   = {123--158},
  year    = {2023}
}

@article{gravell2021learning,
  author  = {Gravell, Benjamin and Esfahani, Peyman Mohajerin and Summers, Tyler},
  title   = {Learning optimal controllers for linear systems with multiplicative noise via policy gradient},
  journal = {IEEE Transactions on Automatic Control},
  volume  = {66},
  number  = {11},
  pages   = {5283--5298},
  year    = {2021}
}

@article{hambly2021policy,
  author  = {Hambly, Ben and Xu, Renyuan and Yang, Huining},
  title   = {Policy gradient methods for the noisy linear quadratic regulator over a finite horizon},
  journal = {SIAM Journal on Control and Optimization},
  volume  = {59},
  number  = {5},
  pages   = {3359--3391},
  year    = {2021}
}

@article{nesterov2017random,
  author  = {Nesterov, Yurii and Spokoiny, Vladimir},
  title   = {Random gradient-free minimization of convex functions},
  journal = {Foundations of Computational Mathematics},
  volume  = {17},
  number  = {2},
  pages   = {527--566},
  year    = {2017}
}

@article{ghadimi2013stochastic,
  author  = {Ghadimi, Saeed and Lan, Guanghui},
  title   = {Stochastic first- and zeroth-order methods for nonconvex stochastic programming},
  journal = {SIAM Journal on Optimization},
  volume  = {23},
  number  = {4},
  pages   = {2341--2368},
  year    = {2013}
}

@article{li2022distributed,
  author  = {Li, Yingying and Tang, Yujie and Zhang, Runyu and Li, Na},
  title   = {Distributed reinforcement learning for decentralized linear quadratic control: A derivative-free policy optimization approach},
  journal = {IEEE Transactions on Automatic Control},
  volume  = {67},
  number  = {12},
  pages   = {6429--6444},
  year    = {2022}
}

@inproceedings{toso2024oracle,
  author    = {Toso, Leonardo F. and Wang, Han and Anderson, James},
  title     = {Oracle complexity reduction for model-free {LQR}: A stochastic variance-reduced policy gradient approach},
  booktitle = {Proceedings of the American Control Conference (ACC)},
  year      = {2024}
}

@inproceedings{mcmahan2017communication,
  author    = {McMahan, H. Brendan and Moore, Eider and Ramage, Daniel and Hampson, Seth and Ag{\"u}era y Arcas, Blaise},
  title     = {Communication-efficient learning of deep networks from decentralized data},
  booktitle = {Proceedings of the 20th International Conference on Artificial Intelligence and Statistics (AISTATS)},
  series    = {PMLR},
  volume    = {54},
  pages     = {1273--1282},
  year      = {2017}
}

@inproceedings{alistarh2017qsgd,
  author    = {Alistarh, Dan and Grubic, Demjan and Li, Jerry and Tomioka, Ryota and Vojnovic, Milan},
  title     = {{QSGD}: Communication-Efficient {SGD} via Gradient Quantization and Encoding},
  booktitle = {Advances in Neural Information Processing Systems},
  volume    = {30},
  pages     = {1709--1720},
  year      = {2017}
}

@article{chen2022optimal,
  author  = {Chen, Wenlin and Horv{\'a}th, Samuel and Richt{\'a}rik, Peter},
  title   = {Optimal client sampling for federated learning},
  journal = {Transactions on Machine Learning Research},
  year    = {2022}
}

@article{neyman1934two,
  author  = {Neyman, Jerzy},
  title   = {On the two different aspects of the representative method: The method of stratified sampling and the method of purposive selection},
  journal = {Journal of the Royal Statistical Society},
  volume  = {97},
  number  = {4},
  pages   = {558--625},
  year    = {1934}
}

@inproceedings{khodadadian2022federated,
  author    = {Khodadadian, Sajad and Sharma, Pranay and Joshi, Gauri and Maguluri, Siva Theja},
  title     = {Federated Reinforcement Learning: Linear Speedup Under {M}arkovian Sampling},
  booktitle = {Proceedings of the 39th International Conference on Machine Learning},
  series    = {Proceedings of Machine Learning Research},
  volume    = {162},
  pages     = {10997--11057},
  year      = {2022}
}

@article{fang2022communication,
  author  = {Fang, Wenzhi and Yu, Ziyi and Jiang, Yuning and Shi, Yuanming and Jones, Colin N. and Zhou, Yong},
  title   = {Communication-Efficient Stochastic Zeroth-Order Optimization for Federated Learning},
  journal = {IEEE Transactions on Signal Processing},
  volume  = {70},
  pages   = {5058--5073},
  year    = {2022}
}

@inproceedings{qin2024federated,
  author    = {Qin, Zhen and Chen, Daoyuan and Qian, Bingchen and Ding, Bolin and Li, Yaliang and Deng, Shuiguang},
  title     = {Federated Full-Parameter Tuning of Billion-Sized Language Models with Communication Cost under 18 Kilobytes},
  booktitle = {Proceedings of the 41st International Conference on Machine Learning},
  series    = {Proceedings of Machine Learning Research},
  volume    = {235},
  pages ={41473--41497},
  year      = {2024}
}

@inproceedings{needell2014stochastic,
  author    = {Needell, Deanna and Ward, Rachel and Srebro, Nathan},
  title     = {Stochastic Gradient Descent, Weighted Sampling, and the Randomized {K}aczmarz Algorithm},
  booktitle = {Advances in Neural Information Processing Systems},
  volume    = {27},
  pages     = {1017--1025},
  year      = {2014}
}

@inproceedings{zhao2015stochastic,
  author    = {Zhao, Peilin and Zhang, Tong},
  title     = {Stochastic Optimization with Importance Sampling for Regularized Loss Minimization},
  booktitle = {Proceedings of the 32nd International Conference on Machine Learning},
  series    = {Proceedings of Machine Learning Research},
  volume    = {37},
  pages     = {1--9},
  year      = {2015}
}

@inproceedings{li2020federated,
  author    = {Li, Tian and Sahu, Anit Kumar and Zaheer, Manzil and Sanjabi, Maziar and Talwalkar, Ameet and Smith, Virginia},
  title     = {Federated Optimization in Heterogeneous Networks},
  booktitle = {Proceedings of Machine Learning and Systems},
  volume    = {2},
  pages     = {429--450},
  year      = {2020}
}

@inproceedings{wang2020tackling,
  author    = {Wang, Jianyu and Liu, Qinghua and Liang, Hao and Joshi, Gauri and Poor, H. Vincent},
  title     = {Tackling the Objective Inconsistency Problem in Heterogeneous Federated Optimization},
  booktitle = {Advances in Neural Information Processing Systems},
  volume    = {33},
  pages     = {7611--7623},
  year      = {2020}
}

@article{rostami2026scalar_federated,
  title={Scalar Communication via Random Direction Refreshing for Distributed Optimization},
  author={Rostami, Mohammadreza and Kia, Solmaz S},
  journal={arXiv preprint arXiv:2610.05666},
  year={2026}
}

@article{rostami2025fedmpdd,
  title={FedMPDD: Communication-Efficient Federated Learning with Privacy Preservation Attributes via Projected Directional Derivative},
  author={Rostami, Mohammadreza and Kia, Solmaz S},
  journal={arXiv preprint arXiv:2512.20814},
  year={2025}
}

@string{an = "The American Naturalist"}

@String{ar = "Autonomous Robots"}

@string{optimization = "Optimization"}

@STRING{acc = "{A}merican {C}ontrol {C}onference"}

@String{SIAM = "SIAM"}
\appendices
\section{Proofs}
\label{app:proofs}

\begin{proof}[Proof of Theorem~\ref{thm:allocation}]
Let $f(m)=\sum_iV_i/m_i$. The feasible set
of~\eqref{eq:relaxed-problem} is compact and convex, and the Hessian
$\diag(2V_i/m_i^3)$ of $f$ is positive definite on it, so the minimizer
$m^\star$ exists and is unique. Every partial derivative of $f$ is
negative, so any allocation with $\sum_im_i<S$ can be improved, and the
budget is exhausted. The KKT conditions
$-V_i/m_i^2+\lambda-\ell_i=0$, $\ell_i\ge0$, $\ell_i(m_i-m_{\min})=0$
give $m_i=\sqrt{V_i/\lambda}$ when the bound is inactive and
$m_i=m_{\min}$ when $\lambda\ge V_i/m_{\min}^2$, which
is~\eqref{eq:optimal-counts}. The map
$\lambda\mapsto\sum_i\max\{m_{\min},\sqrt{V_i/\lambda}\}$ is continuous
and nonincreasing, and strictly decreasing wherever its value exceeds
$Mm_{\min}$; hence the equation $\sum_im_i^\star=S$ has a unique root
when $S>Mm_{\min}$. If no bound is active, summing the counts gives
$\lambda^{-1/2}=S/\sum_j\sqrt{V_j}$ and the proportional form; the
substitution of $m^\star$ into~\eqref{eq:centered-error} gives
$\cV_k(m^\star)$ in~\eqref{eq:variance-optimum}, and the
Cauchy--Schwarz inequality $(\sum_i\sqrt{V_i})^2\le M\sum_iV_i$, with
equality if and only if all $V_i$ coincide, gives the comparison with
$\cV_k(m^{\rm eq})=\sum_iV_i/(MS)$.
\end{proof}

\begin{proof}[Proof of Theorem~\ref{thm:stability}]
On the event $\fro{\Xi_k}\le\tau_k$, every $E$ with $\fro E\le r$
satisfies
$\fro{K_{k+1}+E-K_k}\le\eta_k\fro{\overline G_k}+\eta_k\tau_k+r=d_k$,
so $K_{k+1}+E$ lies in the closed ball of radius $d_k$ about $K_k$,
which is contained in $\cS$. Hence
$\mathsf F_k\subseteq\{\fro{\Xi_k}>\tau_k\}$, and Markov's inequality
applied to $\fro{\Xi_k}^2$ together with~\eqref{eq:centered-error}
gives~\eqref{eq:failure-certificate}.
\end{proof}

\begin{proof}[Proof of Corollary~\ref{cor:stability-gain}]
Item (i) follows from $\cV_k(m^{\rm opt})=\cV_k(m^{\rm eq})/\Gamma_k(S)$.
For item (ii), the certificate $\cV_k(m)\le\delta_k\tau_k^2$ with
$\cV_k(m^\star)$ from~\eqref{eq:variance-optimum} holds if and only if
$S\ge S_k^{\rm per}$, and with $\cV_k(m^{\rm eq})=\sum_iV_{i,k}/(MS)$ if
and only if $S\ge\sum_iV_{i,k}/(M\delta_k\tau_k^2)=\Gamma_kS_k^{\rm per}$.
For item (iii), solving $\cV_k(m)\le\delta_k\tau_k^2$ for $\eta_k$
through~\eqref{eq:stability-tolerance} gives $\eta_k^{\max}(m)$. Put
$a_0=\fro{\overline G_k}$ and
$s=\sqrt{\cV_k(m^{\rm opt})/\delta_k}$, so that
$\sqrt{\cV_k(m^{\rm eq})/\delta_k}=\sqrt{\Gamma_k(S)}\,s$. The
inequalities $a_0+s\le a_0+\sqrt{\Gamma_k(S)}\,s
\le\sqrt{\Gamma_k(S)}\,(a_0+s)$ give the bounds
in~\eqref{eq:certified-step}.
\end{proof}

\begin{proof}[Proof of Corollary~\ref{cor:budget}]
Put $S_+=S-Mm_{\min}$ and $w_i=\sqrt{V_i}/\sum_j\sqrt{V_j}$. The
integer counts $m_i=m_{\min}+\lfloor S_+w_i\rfloor$ satisfy
$\sum_im_i\le S$ and, since $m_{\min}\ge1$, $m_i\ge S_+w_i$; hence
$\sum_iV_i/m_i\le\sum_iV_i/(S_+w_i)=(\sum_i\sqrt{V_i})^2/S_+$. By
Lemma~\ref{lem:greedy} the integer optimum is no worse, and
Proposition~\ref{prop:robust} gives
$\cV_k(\widehat m)\le(1+\epsilon)(\sum_i\sqrt{V_{i,k}})^2/[(1-\epsilon)M^2S_+]$,
which~\eqref{eq:integer-stability-budget} makes at most
$\delta_k\tau_k^2$. The last claim follows from
Proposition~\ref{prop:robust} and
$\cV_k(m^{\rm opt})=\cV_k(m^{\rm eq})/\Gamma_k(S)$.
\end{proof}

\begin{proof}[Proof of Proposition~\ref{prop:scaled}]
Couple each reference rollout with its scaled copy by using the same
direction and the same realization of $w_t^{(0)}$. Zero initialization
and linearity of~\eqref{eq:closed-loop} give
$x_t^{(i)}=\sqrt{\alpha_i}\,x_t^{(0)}$, so the measured costs and the
coefficients~\eqref{eq:scalar-dd} are multiplied by $\alpha_i$, and
$Z_{i,k,s}\overset{d}{=}\alpha_iZ_{0,k,s}$. The scaling of
$\overline G_{i,k}$ and $V_{i,k}$ follows, and with
$\sqrt{V_{i,k}}\propto\alpha_i$ the proportional counts
of~\eqref{eq:variance-optimum} become $S\alpha_i/\sum_j\alpha_j$.
Substituting $V_{i,k}=\alpha_i^2V_{0,k}$ into~\eqref{eq:stability-gain}
and using
$M^{-1}\sum_i\alpha_i^2=\bar\alpha^2+M^{-1}\sum_i(\alpha_i-\bar\alpha)^2$
gives~\eqref{eq:noise-heterogeneity-gain}. For the last claim, by the
coupling, $\widehat G_k\overset{d}{=}\sum_i\sum_{s=1}^{m_i}
\alpha_i(Mm_i)^{-1}Z_{0,k,s}^{(i)}$ with independent reference estimates
$Z_{0,k,s}^{(i)}$; under proportional counts every weight equals
$\alpha_i/(Mm_i)=\bar\alpha/S$, so
$\widehat G_k\overset{d}{=}(\bar\alpha/S)\sum_{j=1}^SZ_{0,k}^{(j)}$,
which is the update of a homogeneous fleet with scale $\bar\alpha$ and
$S$ rollout pairs in total.
\end{proof}

\begin{proof}[Proof of Theorem~\ref{thm:convergence}]
Write $\Delta_k=J(K_k)-J^\star$ and consider the event
$\mathsf E=\{\fro{\Xi_k}\le s_N\text{ for all }k<N\}$. We first show by
induction that on $\mathsf E$ every iterate lies in $\mathcal C$ and
satisfies~\eqref{eq:linear-neighborhood}; this holds at $k=0$. Suppose
it holds up to $k$, and put $g_k=\nabla J(K_k)$ and $e_k=b_k+\Xi_k$, so
that $\fro{e_k}\le\bar b+s_N$ and $K_{k+1}-K_k=-\eta(g_k+e_k)$. Then
$\fro{K_{k+1}-K_k}+r\le\eta(G+\bar b+s_N)+r\le d$, so the segment from
$K_k$ to $K_{k+1}$ and every gain within distance $r$ of $K_{k+1}$ lie
in the ball of radius $d$ about $K_k$, which is contained in $\cS$ and
on which the Hessian of $J$ is bounded by $L$. The descent lemma and
$\eta\le1/L$ give
\begin{align*}
 J(K_{k+1})&\le J(K_k)-\eta\inner{g_k}{g_k+e_k}
 +\frac{L\eta^2}{2}\fro{g_k+e_k}^2\\
 &\le J(K_k)-\frac\eta2\fro{g_k}^2+\frac\eta2\fro{e_k}^2,
\end{align*}
since the cross terms cancel. By~\eqref{eq:aggregate-pl} and
$\mu\le\mu_0$,
$\Delta_{k+1}\le(1-\mu\eta)\Delta_k+\frac\eta2(\bar b+s_N)^2
=(1-\mu\eta)\Delta_k+\mu\eta R_N$. Since $\mu\eta\le1$,
$\Delta_k\le c-J^\star$, and $R_N\le c-J^\star$, this gives
$\Delta_{k+1}\le c-J^\star$, that is, $K_{k+1}\in\mathcal C$; unrolling
the recursion and bounding the geometric sum by $R_N$
gives~\eqref{eq:linear-neighborhood} at $k+1$.

It remains to bound $\Pr(\mathsf E^c)$. Decompose $\mathsf E^c$
according to the first round $k$ at which $\fro{\Xi_k}>s_N$. Before
that round no violation has occurred, so the induction gives
$K_k\in\mathcal C$ and hence $\cV_k(m_k)\le\bar v$, and Markov's
inequality yields $\Pr(\fro{\Xi_k}>s_N\mid\cH_k)\le\bar v/s_N^2=\delta/N$.
Summing over $k<N$ bounds $\Pr(\mathsf E^c)$ by $\delta$; no
independence across rounds is required.
\end{proof}

\begin{proof}[Proof of Corollary~\ref{cor:convergence-gain}]
On histories in $\mathcal C$,
$\cV_k(m^{\rm opt})=\cV_k(m^{\rm eq})/\Gamma_k(S)\le\bar v_{\rm eq}/\Gamma$,
so $\bar v_{\rm per}=\bar v_{\rm eq}/\Gamma$ is admissible and
$s_{N,\rm per}=s_{N,\rm eq}/\sqrt\Gamma$ by definition; under scaled
noise $\Gamma_k(S)$ does not depend on $k$ by
Proposition~\ref{prop:scaled}, and it equals $\Gamma_k$ when the
proportional counts are feasible. Item (i) follows by substitution into
$R_N$. For item (ii), the step ratio follows from
$a_0+s/\sqrt\Gamma\le a_0+s\le\sqrt\Gamma\,(a_0+s/\sqrt\Gamma)$ with
$a_0=G+\bar b$ and $s=s_{N,\rm eq}$. For item (iii), the budget
$S\ge Na/(\delta t_\varepsilon^2)$ and $\cV_k\le a/S$ give
$s_N\le t_\varepsilon$, hence the step condition
in~\eqref{eq:convergence-conditions} and
$R_N\le\varepsilon/2\le c-J^\star$. Since
$N\ge\log(2\Delta_0/\varepsilon)/(\mu\eta)$ gives
$(1-\mu\eta)^N\Delta_0\le e^{-\mu\eta N}\Delta_0\le\varepsilon/2$,
\eqref{eq:linear-neighborhood} yields $\Delta_N\le\varepsilon$. Finally,
$\cV_k(m^{\rm opt})\le\cV_k(m^{\rm eq})/\Gamma$ on $\mathcal C$ implies
that $a$ may be taken $\Gamma$ times smaller for personalized counts.
\end{proof}
\end{document}